\documentclass[twocolumn]{jpsj2} 
\usepackage{times}

\def\runauthor{H. \textsc{Onishi} and T. \textsc{Hotta}}

\title{%
Multipole correlations in low-dimensional
\mbox{\boldmath $f$}-electron systems}

\author{%
Hiroaki \textsc{Onishi} and Takashi \textsc{Hotta}
}

\inst{%
Advanced Science Research Center,
Japan Atomic Energy Agency,
Tokai, Ibaraki 319-1195, Japan
}

\recdate{November 11, 2005}

\abst{%
By using a density matrix renormalization group method,
we investigate the ground-state properties of
a one-dimensional three-orbital Hubbard model
on the basis of a $j$-$j$ coupling scheme.
For $B_4^0$$\ne$0, where $B_4^0$ is a parameter to control
cubic crystalline electric field effect,
one orbital is itinerant, while other two are localized.
Due to the competition between itinerant and localized natures,
we obtain orbital ordering pattern which is sensitive to $B_4^0$,
leading to a characteristic change of $\Gamma_{3g}$ quadrupole state
into an incommensurate structure.
At $B_4^0$=0, all the three orbitals are degenerate, but
we observe a peak at $q$=0 in $\Gamma_{3g}$ quadrupole correlation,
indicating a ferro-orbital state,
and the peak at $q$=$\pi$ in $\Gamma_{4u}$ dipole correlation,
suggesting an antiferromagnetic state.
We also discuss the effect of $\Gamma_{4u}$ octupole
on magnetic anisotropy.
}

\kword{%
multipole correlation,
$j$-$j$ coupling scheme,
density matrix renormalization group method
}

\begin{document}
\maketitle


In recent years, it has been widely recognized that
the orbital degree of freedom is one of the key ingredients
to understand the origin of diverse magnetic properties of
strongly correlated $d$- and $f$-electron systems.\cite{orbital2001}
Since in general, $d$- and $f$-orbitals are spatially anisotropic,
orbital ordering leads to highly non-uniform spin-spin interactions
under the orbital-ordered background,
and there occur various kinds of magnetic structures
according to the type of orbital even in the same lattice structure.
Here we note that for $f$-electron systems, due care should be
paid to treat spin and orbital degrees of freedom,
since spin and orbital are tightly coupled with each other
due to a strong spin-orbit interaction.
In this situation,
total angular momentum provides a good quantum number,
instead of spin and orbital itself.
Such a complex spin-orbital state
can be classified from the viewpoint of \textit{multipole},
which is represented by total angular momentum.

Thus, in $f$-electron systems, ordering of multipole moment
has been discussed intensively both from experimental
and theoretical sides.
Since among multipoles, dipole and quadrupole are corresponding to
ordinary spin and orbital in $d$-electron systems, respectively,
ordering of dipole and/or quadrupole moments
can be grasped by the $d$-electron-like theory.
In this sense, octupole should be the first exotic
multipole moment which we encounter in the solid state physics.
In fact, a possibility of octupole ordering has been actively
discussed for actual $f$-electron materials such as
Ce$_{x}$La$_{1-x}$B$_{6}$,
\cite{Kuramoto2000,Kusunose2001,Kubo2003,Kubo2004,Mannix2005,Kusunose2005}
NpO$_{2}$,\cite{Santini2000,Santini2002,Paixao2002,Caciuffo2003,
Lovesey2003,Kiss2003,Tokunaga2005,Sakai2005,Kubo2005a,Kubo2005b,Kubo2005c}
and SmRu$_4$P$_{12}$.\cite{Yoshizawa2005,Hachitani2005}

In order to gain deep insight into multipole phenomena
of $f$-electron systems beyond the phenomenological level,
it is quite important to clarify multipole properties
of a microscopic model for $f$-electron materials.
However, this is a rather difficult task due to the complexity
of the $f$-electron model which should include
charge, spin, and orbital degrees of freedom simultaneously.
For the purpose, the construction of a microscopic model for
$f$-electron systems based on the $j$-$j$ coupling scheme
has been proposed.\cite{Hotta2003}
By analyzing such an $f$-electron model, for instance,
the microscopic origin of octupole ordering in NpO$_2$ has
recently been established.\cite{Kubo2005a,Kubo2005b,Kubo2005c}
Furthermore, magnetism and superconductivity of several kinds
of $f$-electron materials have been studied from a microscopic viewpoint.
\cite{Takimoto2002,Takimoto2003,Takimoto2004,Hotta2004a,Hotta2004b,
Onishi2004,Hotta2005a,Hotta2005b,Hotta2005c}

However, multipole properties of the periodic systems
have not been investigated satisfactorily at the microscopic level,
even though their potential role has been emphasized
in several kinds of $f$-electron materials.
In the $j$-$j$ coupling scheme, we can microscopically evaluate
physical quantities related to multipoles by treating them
as combined spin and orbital degrees of freedom.
In this paper, thus, we develop a numerical technique to study
multipole properties from a microscopic viewpoint
on the basis of the $j$-$j$ coupling scheme.
We report detailed microscopic analysis
about multipole properties in a one-dimensional model
for $f$-electron systems.
It is found that incommensurability of
quadrupole correlation function is changed by the effect of
crystalline electric field (CEF).
We also observe the change of magnetic anisotropy due to
the effect of octupole magnetic moment which belong to
the same symmetry of dipole moment.
Throughout this paper, we use such units as $\hbar$=$k_{\rm B}$=1.


First let us explain briefly the model construction
for $f$-electron systems.\cite{Hotta2003}
In the $j$-$j$ coupling scheme,
we first include the strong spin-orbit interaction,
and consider only the lower sextet with $j$=5/2(=3$-$1/2),
where $j$ is the total angular momentum.
Taking the effects of CEF, Coulomb interaction,
and $f$-electron hopping into consideration,
we can obtain a multi-orbital Hubbard model
for $f$-electron systems.
In this paper, we study the ground-state properties of
a one-dimensional system along the $z$ direction with $N$ sites
including one electron per site under the cubic CEF effect.
We note that due to the cibic symmetry, the results do not
depend on the chain direction.

The Hamiltonian is, then, given by
\begin{align}
 H=
 & \sum_{i,\mu,\nu}
   (t_{\mu\nu}^z f_{i\mu}^{\dag}f_{i+1\nu}+\text{h.c.})
   +\sum_{i,\mu,\nu} B_{\mu\nu} f_{i\mu}^{\dag}f_{i\nu}
 \nonumber\\
 & +(1/2)\sum_{i,\mu_1\sim\mu_4}
   I_{\mu_1,\mu_2,\mu_3,\mu_4}
   f_{i\mu_1}^{\dag}f_{i\mu_2}^{\dag}f_{i\mu_3}f_{i\mu_4},
\end{align}
where $f_{i\mu}$ is the annihilation operator for an $f$ electron
with the $z$ component of the total angular momentum $\mu$ at site $i$.
The hopping amplitudes are estimated from the overlap integrals
between $f$-electron wavefunctions, which are given by
$t_{\pm 1/2,\pm 1/2}^z$=8$t_0$ with $t_0$=$(3/56)(ff\sigma)$,
and zero for other cases.
Here $(ff\sigma)$ is $f$-electron hopping amplitude through
the sigma bond between nearest neighbor sites.
Hereafter, we take $t_0$=1 as the energy unit.

In the cubic CEF term, $B_{\mu\nu}$ is expressed by
a cubic CEF parameter, $B_4^0$.
Note that due to the cubic CEF effect,
the sextet with $j$=5/2 is split into
$\Gamma_7$ doublet and $\Gamma_8$ quartet states,
which are represented by
\begin{align}
 |\Gamma_{7\pm}\rangle &=
 \sqrt{1/6}|\pm 5/2\rangle-\sqrt{5/6}|\mp 3/2\rangle,
 \\
 |\Gamma_{8\pm}^a\rangle &=
 \sqrt{5/6}|\pm 5/2\rangle+\sqrt{1/6}|\mp 3/2\rangle,
 \\
 |\Gamma_{8\pm}^b\rangle &=
 |\pm 1/2\rangle,
\end{align}
where $\Gamma_8^a$ and $\Gamma_8^b$ orbitals are introduced
to distinguish two Kramers doublets in $\Gamma_8$,
the subscript $\pm$ denotes pseudospin to label two states
in each Kramers doublet,
and $|j_z\rangle$ in the right hand side is the eigenstate of
the $z$ component of the total angular momentum $j_z$.
The Coulomb integrals $I$ in the $j$-$j$ coupling scheme are expressed
by three Racah parameters, $E_0$, $E_1$, and $E_2$.\cite{Hotta2003}

We investigate the $f$-electron model (1) numerically by exploiting
a density matrix renormalization group (DMRG) method,\cite{White1992}
which has been developed for the numerical analysis of
low-dimensional strongly correlated $d$-electron systems
and quantum spin systems.
In the present calculations,
the finite-system algorithm is used for $N$=16 chains
with the open boundary condition.
The number of states kept for each block $m$ is up to $m$=300,
and the truncation error is estimated to be $10^{-6}$ at most.
Note that the number of bases is 64 for one site due to the three orbitals,
indicating that the size of the superblock Hilbert space becomes
very large as $m^2$$\times$$64^2$.
To reduce the size of the Hilbert space, in general,
it is useful to decompose the Hilbert space into a block-diagonal form
by using symmetries of the Hamiltonian.
We mention here that the Coulomb interaction is diagonal
in terms of $J_z^{\text{tot}}$, which is the total of $j_z$,
but the diagonal blocks classified by $J_z^{\text{tot}}$ are mixed
due to the hopping and CEF terms.
In fact, we can make use of the integer $2J_z^{\text{tot}}\pmod{8}$
as a good quantum number,
as well as the total number of electrons.
We emphasize that this is the first trial to apply the DMRG
method to the microscopic $f$-electron model
on the basis of the $j$-$j$ coupling scheme.


\begin{figure}[t]
\begin{center}
\includegraphics[width=0.48\textwidth]{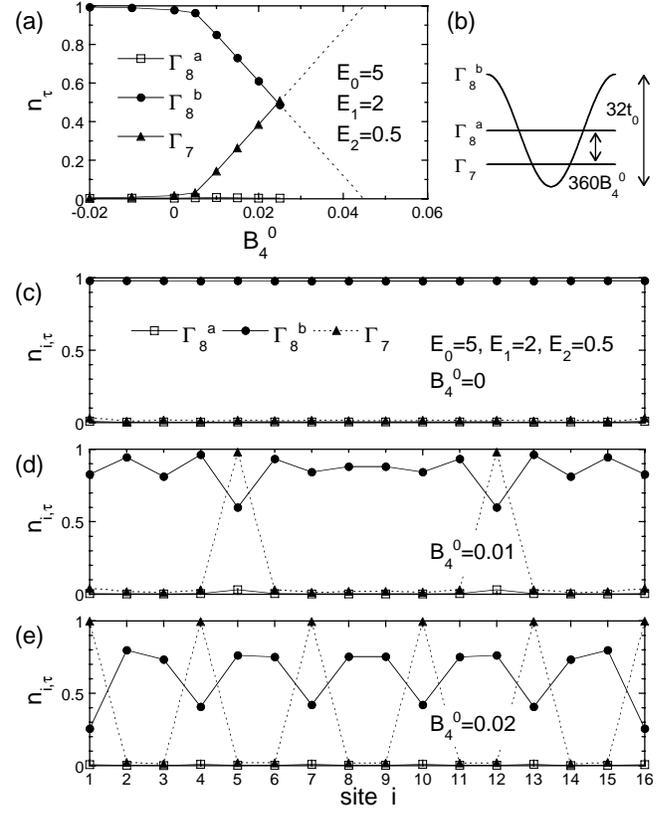}
\end{center}
\caption{%
(a) Electron densities
in the $\Gamma_8^a$, $\Gamma_8^b$, and $\Gamma_7$ orbitals.
The dotted lines denote the linear extrapolation
to $n_{\Gamma_7}$=1 and $n_{\Gamma_8^b}$=0.
(b) Schematic view of the free-electron energy dispersion.
Local electron densities
for (c) $B_4^0$=0, (d) $B_4^0$=0.01, and (e) $B_4^0$=0.02.
}
\end{figure}

Let us first discuss the change of the orbital state
due to the CEF effect.
In Fig.~1(a),
we show the $B_4^0$ dependence of the electron densities
$n_{\tau}$=%
$\sum_{i,\sigma}\langle f_{i\tau\sigma}^{\dag}f_{i\tau\sigma} \rangle/N$
in the $\Gamma_8^a$, $\Gamma_8^b$, and $\Gamma_7$ orbitals
for $E_0$=5, $E_1$=2, and $E_2$=0.5,
where $f_{i\tau\sigma}$ is the annihilation operator for an $f$ electron
with pseudospin $\sigma$ in orbital $\tau$ at site $i$.
At $B_4^0$=0,
all the three orbitals are degenerate,
but we find that the $\Gamma_8^b$ orbital is favorably occupied,
while the $\Gamma_8^a$ and $\Gamma_7$ orbitals are almost empty.
This result can be intuitively understood
from the free-electron energy dispersion,
composed of a cosine band due to itinerant $\Gamma_8^b$ orbitals
and flat bands due to localized $\Gamma_8^a$ and $\Gamma_7$ orbitals,
as shown in Fig.~1(b).
At $B_4^0$=0, where two flat bands are degenerate,
the $\Gamma_8^b$ cosine band is occupied
from the bottom up to the flat-band level,
in the case of one electron per site.
Namely, it is favorable to occupy itinerant $\Gamma_8^b$ orbitals
to gain kinetic energy by electron hopping.
For $B_4^0$$<$0, electrons occupy the lower $\Gamma_8$ level,
and we find no remarkable change in the orbital densities.
On the other hand, for $B_4^0$$>$0,
electrons are forced to occupy the lower $\Gamma_7$ level
to gain CEF potential energy.
Note that in the free-electron energy dispersion,
the $\Gamma_7$ flat band goes down
to the bottom of the $\Gamma_8^b$ cosine band
at $360B_4^0$=$16t_0$, i.e., $B_4^0/t_0$=0.044.
If we estimate the value of $B_4^0/t_0$ that satisfies $n_{\Gamma_7}$=1
from the present DMRG results, we obtain $B_4^0/t_0$$\sim$0.045,
which is consistent with the estimation for the free-electron case.
In order to obtain convergent results for $B_4^0$$\ge$0.03,
it is necessary to enlarge $m$ and/or $N$ in the DMRG calculations.
Unfortunately, we have not yet finished such DMRG calculations,
mainly due to the limitation of CPU time and memory resources.
In near future, we obtain DMRG results also for $B_4^0$$\ge$0.03,
which will be shown elsewhere.

Within the present DMRG calculations,
let us discuss the characteristics of the change
in the orbital structure due to the CEF effect.
For the purpose, we investigate the local electron densities
$n_{i\tau}$=%
$\sum_{\sigma}\langle f_{i\tau\sigma}^{\dag}f_{i\tau\sigma} \rangle$.
As shown in Fig.~1(c),
$\Gamma_8^b$ orbitals are equally occupied in every site at $B_4^0$=0.
On the other hand,
when $B_4^0$ is increased,
we observe that electrons are accommodated in $\Gamma_7$ orbitals
in some sites,
in which the electron density in the $\Gamma_7$ orbital is practically one,
as shown in Figs.~1(d) and 1(e).
Moreover, such $\Gamma_7$ sites are arranged at regular intervals,
indroducing cuts of the chain
due to the absence of electron hopping,
while the $\Gamma_8^b$ sites form clusters
to gain kinetic energy in the clusters.
Note that the clusters of the $\Gamma_8^b$ sites should weakly couple
with each other through the $\Gamma_7$ sites,
since the $\Gamma_8^b$ orbital is also occupied
in the $\Gamma_7$ sites to some extent.
Thus, the incommensurate orbital ordering pattern
appears due to the competition
between the itinerant $\Gamma_8$ and localized $\Gamma_7$ orbitals.

Now we turn our attention to multipole properties
to discuss the ground-state properties from the viewpoint of multipole.
In order to clarify what types of multipole correlations develop,
we evaluate the multipole correlation function
\begin{equation}
\chi_{\Gamma_{\gamma}}(q)=
(1/N)\sum_{k,l}e^{iq(k-l)}
\langle X_{k\Gamma_{\gamma}}X_{l\Gamma_{\gamma}} \rangle,
\end{equation}
where
$X_{i\Gamma_{\gamma}}$ is the multipole operator
with the symbol $X$ of multipole
for the irreducible representation $\Gamma_{\gamma}$
in the cubic symmetry at site $i$,
and $\langle\cdots\rangle$ denotes the average
using the ground-state wavefunction.
The definition of the multipole operators is given in Table~I.
Note that we consider the multipole operators up to rank 3,
and there are 15 types of multipoles including
three dipoles, five quadrupoles, and seven octupoles.\cite{Shiina1997}

\begin{table}[t]
\begin{center}
\begin{tabular}{ll}
\hline
$\Gamma_{\gamma}$ multipole & multipole operator \\
\hline
$\Gamma_{4u}$ dipole
& $J_{4ux}$=$J_x$, $J_{4uy}$=$J_y$, $J_{4uz}$=$J_z$
\\
$\Gamma_{3g}$ quadrupole
& $O_{3gu}$=(1/2)(2$J_z^2$$-$$J_x^2$$-$$J_y^2$)
\\
& $O_{3gv}$=($\sqrt{3}$/2)($J_x^2$$-$$J_y^2$)
\\
$\Gamma_{5g}$ quadrupole
& $O_{5gx}$=($\sqrt{3}$/2)$\overline{J_y J_z}$
\\
& $O_{5gy}$=($\sqrt{3}$/2)$\overline{J_z J_x}$
\\
& $O_{5gz}$=($\sqrt{3}$/2)$\overline{J_x J_y}$
\\
$\Gamma_{2u}$ octupole
& $T_{2u}$=($\sqrt{15}$/6)$\overline{J_x J_y J_z}$
\\
$\Gamma_{4u}$ octupole
& $T_{4ux}$=(1/2)(2$J_x^3$$-$$\overline{J_x J_y^2}$$-$$\overline{J_z^2 J_x}$)
\\
& $T_{4uy}$=(1/2)(2$J_y^3$$-$$\overline{J_y J_z^2}$$-$$\overline{J_x^2 J_y}$)
\\
& $T_{4uz}$=(1/2)(2$J_z^3$$-$$\overline{J_z J_x^2}$$-$$\overline{J_y^2 J_z}$)
\\
$\Gamma_{5u}$ octupole
& $T_{5ux}$=($\sqrt{15}$/6)($\overline{J_x J_y^2}$$-$$\overline{J_z^2 J_x}$)
\\
& $T_{5uy}$=($\sqrt{15}$/6)($\overline{J_y J_z^2}$$-$$\overline{J_x^2 J_y}$)
\\
& $T_{5uz}$=($\sqrt{15}$/6)($\overline{J_z J_x^2}$$-$$\overline{J_y^2 J_z}$)
\\
\hline
\end{tabular}
\end{center}
\caption{%
Multipole operators up to rank 3.
The overline on the product denotes the operation of taking
all possible permutations in terms of cartesian components,
e.g., $\overline{J_x J_y}$=$J_x J_y$+$J_y J_x$.
}
\end{table}

\begin{figure}[t]
\begin{center}
\includegraphics[width=0.48\textwidth]{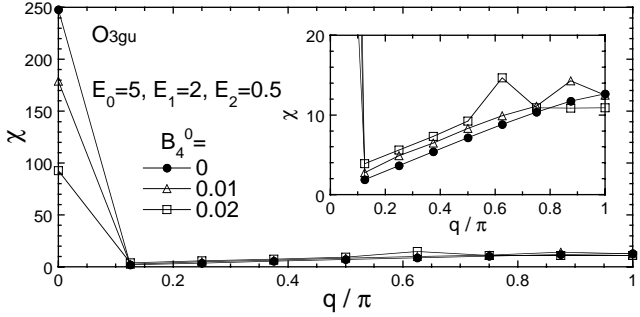}
\end{center}
\caption{%
$O_{3gu}$ quadrupole correlation functions
for $B_4^0$=0, 0.01, and 0.02.
Inset shows the results in the region of small $\chi$
in a magnified scale.
}
\end{figure}

\begin{figure}[t]
\begin{center}
\includegraphics[width=0.48\textwidth]{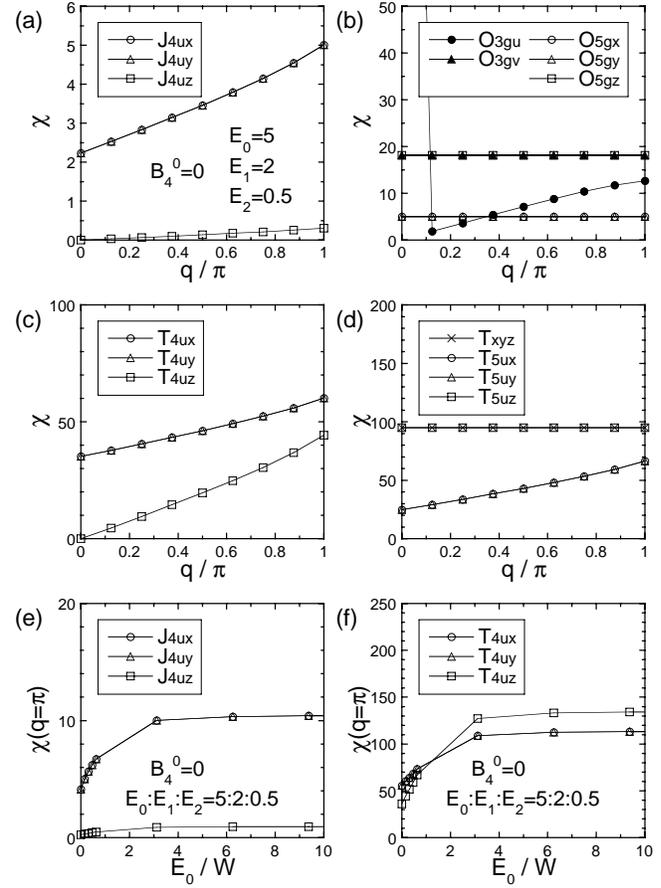}
\end{center}
\caption{%
Multipole correlation functions of
(a) $\Gamma_{4u}$ dipoles,
(b) $\Gamma_{3g}$ and $\Gamma_{5g}$ quadrupoles,
(c) $\Gamma_{4u}$ octupoles,
and (d) $\Gamma_{2u}$ and $\Gamma_{5u}$ octupoles
at $B_4^0$=0 for $E_0$=5, $E_1$=2 and $E_2$=0.5.
The values at the peak $q$=$\pi$ of the correlation functions of
the magnetic $\Gamma_{4u}$ (e) dipole and (f) octupole at $B_4^0$=0.
The ratio $E_0$:$E_1$:$E_2$ is fixed to 5:2:0.5
and the magnitude is varied.
In the horizontal axis,
$E_0$ is normalized by the bandwidth $W$=$32t_0$.
}
\end{figure}

In Fig.~2, among 15 types of multipoles,
we show a typical result of the $O_{3gu}$ quadrupole correlation function
for several values of $B_4^0$.
A sharp main peak at $q$=0 is observed,
indicating a $\Gamma_8^b$ ferro-orbital state,
and its amplitude is reduced with increasing $B_4^0$,
since the electron density in the $\Gamma_8^b$ orbital is suppressed
due to the CEF effect.
On the other hand, as shown in the inset of Fig.~2,
we observe a small peak at $q$=$\pi$ for $B_4^0$=0,
while the position of the peak moves away from $q$=$\pi$
to an incommensurate one with the increase of $B_4^0$.
This characteristic change in the incommensurability
of the $\Gamma_{3g}$ quadrupole state due to the CEF effect
is in accordance with the change of the orbital structure.

Let us here consider multipole properties
also for the other components in more detail,
with particular attention to the case of $B_4^0$=0.
In Figs.~3(a)-(d),
we show the multipole correlation functions.
As shown in Fig.~3(a),
we observe a peak at $q$=$\pi$
in each of the $\Gamma_{4u}$ dipole correlation functions,
suggesting an antiferromagnetic state.
On the other hand, as shown in Fig.~3(b),
the $O_{3gu}$ quadrupole correlation function has a peak at $q$=0
with large amplidute beyond the range of this figure
(see also Fig. 2),
indicating a $\Gamma_8^b$ ferro-orbital state,
as intensively discussed above.
Note that the correlation functions
of the quadrupole moments other than $O_{3gu}$
become flat without any significant structure,
since they behave independently at each site.
Note also that the correlation functions of $O_{3gv}$ and $O_{5gz}$
coincide with each other in spite of their different symmetries,
since in the atomic limit, the expectation values
$\langle O_{3gv}^2 \rangle$ and $\langle O_{5gz}^2 \rangle$
taken by the $\Gamma_8^b$ state are found to be equivalent.
In Figs.~3(c) and 3(d), we show
octupole correlation functions.
As observed in Fig.~3(c),
the $\Gamma_{4u}$ octupole correlation functions
have a peak at $q$=$\pi$,
in a way similar to the case for $\Gamma_{4u}$ dipole,
as observed in Fig.~3(a).
In Fig.~3(d), we also find a peak at $q$=$\pi$
in the $T_{5ux}$ and $T_{5uy}$ correlation functions.
Note that the $T_{2u}$ and $T_{5uz}$ correlation functions
become flat and they agree well with each other, since again,
$\langle T_{2u}^2 \rangle$=$\langle T_{5uz}^2 \rangle$
in the atomic limit assuming that the ground state is
$\Gamma_8^b$.

Here we remark the anisotropic behavior
of the magnetic $\Gamma_{4u}$ dipole and octupole moments,
which belong to the same symmetry.
In Fig.~3(a), we find that
the correlation function of $J_{4uz}$ dipole moment is suppressed
in comparison with those of $J_{4ux}$ and $J_{4uy}$.
This anisotropy is naturally understood from the fact that
$\Gamma_{8\pm}^b$ states are described by
$|$$\pm 1/2\rangle$ among $j$=5/2 multiplets.\cite{Hotta2004b}
Namely, dipole moment lies in the $x$-$y$ plane at a local level
due to the $\Gamma_8^b$ ferro-orbital state,
and consequently,
the correlation in the $x$-$y$ plane exhibits a larger value
than that along the $z$ direction.
On the other hand, as shown in Fig.~3(c),
the correlation function of $T_{4uz}$ octupole moment is also
reduced from those of $T_{4ux}$ and $T_{4uy}$,
which is the same behavior in the $\Gamma_{4u}$ dipole.

Let us also discuss the magnetic anisotropy
in the strong-coupling limit.
In Figs.~3(e) and 3(f),
we show the values at the peak $q$=$\pi$ of
the $\Gamma_{4u}$ dipole and octupole correlation functions,
respectively,
as a function of the magnitude of the Coulomb interaction
with the ratio $E_0$:$E_1$:$E_2$ fixed to 5:2:0.5.
It is observed in common that the correlation is enhanced
due to the Coulomb interaction,
but we find distinctive behavior in the anisotropy
between dipole and octupole.
In Fig.~3(e), we find that
the $J_{4uz}$ dipole correlation remains smaller than
those of $J_{4ux}$ and $J_{4uy}$,
even when the Coulomb interaction is increased,
since the ground state is basically described by
the $\Gamma_8^b$ ferro-orbital state still in the strong-coupling region.
On the other hand, as shown in Fig.~3(f),
the $T_{4uz}$ octupole correlation is much enhanced
compared with those of $T_{4ux}$ and $T_{4uy}$,
and changes to be dominant in the strong-coupling region.
Note that this anisotropy cannot be explained
by analogy of the anisotropy of the expectation values
of octupole moments in the atomic limit, since
$\langle T_{4ux}^2 \rangle$=$\langle T_{4uy}^2 \rangle$%
$>$$\langle T_{4uz}^2 \rangle$ for the $\Gamma_{8}^b$ state.
Namely, this anisotropy could be attributed to anisotropic interactions
between $\Gamma_{4u}$ octupole moments in the strong-coupling region.
\cite{Kubo2005c}
Thus, the change of magnetic anisotropy appears
in the strong-coupling region
due to the effect of the $\Gamma_{4u}$ octupole moments.

Finally, we note that with increasing $B_4^0$,
all the multipole correlations are suppressed.
Since electrons in the $\Gamma_7$ orbital cannot move in the chain,
multipole moments fluctuate independently at each site.
On the other hand, we can observe incommensurate structures
in the multipole correlation functions,
as shown in Fig.~2.
The detail of the $B_4^0$ dependence of the multipole correlations
will be discussed elsewhere in future.


In summary,
we have studied multipole properties of $f$-electron systems
on the basis of the three-orbital Hubbard model,
by using the DMRG method.
At $B_4^0$=0,
there appears the antiferromagnetic/ferro-orbital state,
since itinerant $\Gamma_8^b$ orbitals are occupied to gain kinetic energy.
With increasing $B_4^0$,
electrons are accommodated in localized $\Gamma_7$ orbitals,
leading to the incommensurate orbital structure.
We have also found the charactersitc behavior in magnetic anisotropy
of the $\Gamma_{4u}$ dipole and octupole.


We thank K. Kubo and K. Ueda for discussions.
T.H. is supported by the Japan Society for the Promotion of Science
and by the Ministry of Education, Culture, Sports, Science,
and Technology of Japan.


\end{document}